\documentclass[12pt,twoside]{article}

\usepackage[backend=bibtex,style=numeric-comp,autocite=plain,sorting=none,%
giveninits=true]{biblatex}
\usepackage{amsmath,amsfonts,amssymb,amsthm,mathabx}
\usepackage{times}
\usepackage{hyperref}
\hypersetup{
  breaklinks=true,   
  colorlinks=true,   
  pdfusetitle=true,  
}
\usepackage{xurl}
\usepackage{tensor}
\usepackage{color}
\usepackage{caption}
\usepackage{enumitem}
\usepackage{graphicx}
\usepackage{array}
\usepackage{amsbsy}
\usepackage{appendix}
\usepackage{mathrsfs}
\usepackage{physics}
\usepackage{cancel}
\usepackage{yfonts}
\usepackage{inputenc}
\usepackage{chngcntr}
\usepackage{xcolor}
\usepackage{soul}
\usepackage{cancel}
\usepackage[leqno,fleqn,intlimits,newmultline]{empheq}

\advance\textheight\topskip
\numberwithin{equation}{section} \makeatletter
\@addtoreset{equation}{section}

\begin{document}

\title{Geometric action for extended Bondi-Metzner-Sachs group in four dimensions}

\author{Glenn Barnich, Kevin Nguyen, Romain Ruzziconi}


\def\mytitle{Geometric action for extended Bondi-Metzner-Sachs group in four dimensions}

\pagestyle{myheadings} \markboth{\textsc{\small G.~Barnich, K. Nguyen, R. Ruzziconi}}
{\textsc{\small Geometric action for BMS$_4$}}

\addtolength{\headsep}{4pt}

\begin{centering}

  \vspace{1cm}

  \textbf{\Large{\mytitle}}

    \vspace{1.5cm}

    {\large Glenn Barnich$^a$, Kevin Nguyen$^b$, Romain Ruzziconi$^c$}

\vspace{1cm}

\begin{minipage}{.9\textwidth}\small \it \begin{center}
    $^a$ Physique Th\'eorique et Math\'ematique \\ Universit\'e libre de
    Bruxelles and International Solvay Institutes\\ Campus Plaine C.P. 231,
    B-1050
    Bruxelles, Belgium \\
    E-mail: \href{mailto:Glenn.Barnich@ulb.be}{Glenn.Barnich@ulb.be}
\end{center}
\end{minipage}

\vspace{.5cm}

\begin{minipage}{.9\textwidth}\small \it \begin{center}
    $^b$ Department of Mathematics, King’s College London\\
    The Strand, London WC2R 2LS, UK \\
    E-mail: \href{mailto:kevin.nguyen@kcl.ac.uk}{kevin.nguyen@kcl.ac.uk}
  \end{center}
\end{minipage}

\vspace{.5cm}

\begin{minipage}{.9\textwidth}\small \it \begin{center}
    $^c$ Institute for Theoretical Physics, TU Wien\\
    Wiedner Hauptstrasse 8, A-1040 Vienna, Austria\\
    E-mail: \href{mailto:romain.ruzziconi@tuwien.ac.at}{romain.ruzziconi@tuwien.ac.at}
  \end{center}
\end{minipage}

\end{centering}

\vspace{1cm}
  
\begin{center}
  \begin{minipage}{.9\textwidth} \textsc{Abstract}. The constrained Hamiltonian
    analysis of geometric actions is worked out before applying the construction
    to the extended Bondi-Metzner-Sachs group in four dimensions. For any
    Hamiltonian associated with an extended BMS$_4$ generator, this action
    provides a field theory in two plus one spacetime dimensions whose Poisson
    bracket algebra of Noether charges realizes the extended BMS$_4$ Lie
    algebra. The Poisson structure of the model includes the classical version
    of the operator product expansions that have appeared in the context of
    celestial holography. Furthermore, the model reproduces the evolution
    equations of non-radiative asymptotically flat spacetimes at null infinity.
 \end{minipage}
\end{center}

\vfill
\thispagestyle{empty}

\newpage

\tableofcontents

\vfill
\newpage

\section{Introduction}
\label{sec:introduction}

Whereas sigma models on symmetric spaces $G/H$ involve a Killing form for the
construction of a $G$-invariant spacetime action principle, geometric actions
yield $G$-invariant first order worldline action principles for
$G/\mathcal H_{b_0}$ that do not rely on an invariant metric but rather on the
coadjoint representation and the choice of a fixed coadjoint vector $b_0$ with
$\mathcal H_{b_0}$ the stabilizer subgroup of $b_0$. Geometric actions appear in
the context of the orbit method
\cite{Souriau1970,Kostant1970,kirillov1976elements,A.A.Kirillov897} when
constructing group characters through path integral quantization
\cite{Alekseev:1988vx}. For infinite-dimensional groups such as Kac-Moody or
Virasoro groups \cite{Alekseev:1988ce,Alekseev:1990mp} (see also
e.g.~\cite{Rai:1989js,Delius1990,Aratyn:1990dj}), they appear as Hamiltonian
gauge field theories with spatial sections that are circles. In the same
context, a geometric action for the (centrally extended) BMS group in three
spacetime dimensions \cite{Ashtekar1997,Barnich:2006avcorr} has recently been
constructed in \cite{Barnich:2017jgw}.

The main objective of the current paper is to apply this construction to the BMS
group in four spacetime dimensions
\cite{Bondi:1962px,Sachs1962a,Sachs1962,Newman1966}, or more precisely the
extended version \cite{Barnich:2009se,Barnich:2010eb} that appears in the
celestial holography program \cite{Strominger:2013jfa,Kapec:2014opa,%
  Cachazo:2014fwa,Pasterski:2016qvg,Strominger:2017zoo}, by combining the
ingredients of the construction in three dimensions with the detailed
understanding of the coadjoint representation in four dimensions
\cite{Barnich:2021dta}.

In the course of the construction, the question whether one may consider the
coadjoint vectors as time-dependent, dynamical variables in addition to the
group elements has come up. In order to clarify this issue, after a brief review
of geometric actions, we work out the constrained Hamiltonian analysis for
geometric actions (see also \cite{Rebolledo2015}), before applying the
construction to the group of interest.

In the final two sections, general comments about the relevance of the models to
non-radiative asymptotically flat gravity at null infinity are provided. In
particular, the proper choice of a Hamiltonian reproduces the time-dependence of
the asymptotic symmetry vectors and gravitational flux-balance relations,
without the fluxes.

\section{Generalities}
\label{sec:generalities}

We briefly summarize here the construction of geometric actions following the
conventions of \cite{Barnich:2017jgw}, up to an overall sign in the definition
of the Hamiltonian and the Noether charges. Furthermore, the Lie algebra bracket 
for the diffeomorphism group in 1 dimension is taken here to be minus the Lie bracket
of vector fields, as it should if all signs related to the passage from the group
to the algebra are to be correct. As a consequence, some formulas will have
the opposite signs to those that appear commonly in the literature. 

Let $g\in G$ be a group element and $b_0\in \mathfrak{g}^*$ a fixed coadjoint
vector. The coadjoint orbit $\mathcal{O}_{b_0}$ is the set of coadjoint vectors
$b={\rm Ad}^*_{g^{-1}}b_0 \in \mathfrak{g}^*$ that can be reached from $b_0$
through the coadjoint action ${\rm Ad}^*$. This orbit is isomorphic to
$G/\mathcal{H}_{b_0}$ with $\mathcal{H}_{b_0}$ the isotropy sub-group of $b_0$,
i.e., the elements $h\in G$ such that ${\rm Ad}^*_hb_0=b_0$.

Consider left/right translations by elements $g$,
\begin{equation}
  \label{eq:16}
  L_g: h\to gh\quad / \quad R_g: h\to hg,
\end{equation}
and let $\theta/\kappa$ be the left-invariant/right-invariant Maurer-Cartan
forms satisfying
\begin{equation}
  \label{eq:4}
  d\theta=-\frac 12 {\rm ad}_\theta\theta\quad /\quad d\kappa=\frac 12 {\rm
    ad}_\kappa\kappa.
\end{equation}
For $X\in\mathfrak{g}=\frac{d}{ds}g(s)|_{s=0}$, 
let
\begin{equation}
  \label{eq:5}
  v^R_X=\frac{d}{ds}\big(h g(s)\big)\big|_{s=0}\quad/ \quad
  v^L_X=\frac{d}{ds}\big(g(s) h\big)\big|_{s=0},
\end{equation}
the vector fields which generate right/left translations. These vector
fields are the left/right invariant vector fields that reduce to $X$
at the identity, with
\begin{equation}
  \label{eq:48}
  i_{v^R_X}\theta= X\quad / \quad i_{v^L_X}\kappa= X,
\end{equation}
and 
\begin{equation}
  \label{eq:8}
  i_{v^L_X}\theta={\rm Ad}_{g^{-1}}X\quad
  /\quad i_{v^R_X}\kappa={\rm Ad}_{g}X. 
\end{equation}
On the level of the generators, the left/right invariance of the Maurer-Cartan
forms translates into 
\begin{equation}
  \label{eq:9}
  \pounds_{v^L_X}\theta=0\quad/ \quad \pounds_{v^R_X}\kappa =0. 
\end{equation}

The presymplectic potential and two-form are
\begin{equation}
  \label{eq:2}
  a=\langle b,\theta \rangle= \langle b_0,\kappa \rangle,\quad \Omega= da,
\end{equation}
with $\langle \cdot,\cdot \rangle$ the pairing between $\mathfrak{g}^*$ and
$\mathfrak{g}$. Furthermore, let $X_0\in \mathfrak{g}$ be a fixed Lie algebra
element and
\begin{equation}
  \label{eq:3}
  H_{X_0}=\langle b,X_0 \rangle.
\end{equation}
In particular, it follows from the second of \eqref{eq:8} that
\begin{equation}
  \label{eq:15}
  dH_{X_0}=-i_{v^R_{X_0}}\Omega.
\end{equation}

The geometric actions we will be considering below are of the form
\begin{equation}
  \label{eq:1}
  I_G[g;b_0,X_0]=\int_\gamma \big[a-H_{X_0}dt\big]=\int dt\
  \mathcal{L}_H, \quad
  \mathcal{L}_H=i_Va -H_{X_0},
\end{equation}
where $\gamma: t\to g(t)$ represents a path on $G$ with tangent vector $V=\dot
g$. Models with different $b_0$'s that belong to the same coadjoint orbit are
equivalent in the sense that they are described by geometric actions that are
related by ``field redefinitions'', that is to say invertible reparametrizations
of the configuration space variables that correspond to left translations,
\begin{equation}
  \label{eq:56}
  I_G[g;{\rm Ad}^*_{h^{-1}}b_0,X_0]=I_G[g';b_0,X_0],\quad g'=hg,
\end{equation}
for constant $h\in G$.

In other words, in order to cover all inequivalent models of this type, it is
enough to study them for the different representatives $b_0$ of the partition of
$\mathfrak g^*$ into coadjoint orbits.

For a family $g(t,\lambda)$ of such paths, and their infinitesimal variation
characterized by $W=\frac{\partial g}{\partial \lambda}$, the associated
variation of the action is
\begin{equation}
  \label{eq:6}
  \delta S=\delta\lambda\int dt\big[i_W(-i_V\Omega-d H_{X_0})+\frac{d}{dt} (i_W a)\big],
\end{equation}
so that extremal paths satisfy the equations of motion
\begin{equation}
  \label{eq:7}
  i_V\Omega+dH_{X_0}=0 \iff \langle  b_0,[i_V\kappa -{\rm Ad}_gX_0,\kappa] \rangle=0. 
\end{equation}

For a time dependent Lie algebra element $X=X(t)\in \mathfrak g$, let 
\begin{equation}
  \label{eq:11}
  Q_{X}=\langle  b, X \rangle.
\end{equation}
Under an infinitesimal right translation generated by $v^R_X$, the variation of
the Lagrangian density is
\begin{equation}
  \label{eq:10}
  \frac{d}{dt}Q_X-i_V d Q_X-Q_{[X,X_0]}=Q_{\dot X}-Q_{[X,X_0]}.
\end{equation}
It follows that right translations define global symmetries
if the time dependence of $X(t)$ is fixed through
\begin{equation}
  \label{eq:14}
  \dot X=-{\rm ad_{X_0} X}=[X,X_0]. 
\end{equation}
The associated Noether charges are $Q_X$. When acting with a global symmetry,
they satisfy
\begin{equation}
  \label{eq:12}
  \pounds_{v^R_{X_1}}Q_{X_2}=Q_{[X_1,X_2]}. 
\end{equation}

The little algebra $\mathfrak{h}_{b_0}$ is the subalgebra defined by elements
$\epsilon \in \mathfrak{g}$ such that ${\rm ad}^*_{\epsilon} b_0=0$. As will be
explicitly shown below, the zero eigenvectors of $\Omega$ are exhausted by the
vector fields $v^L_\epsilon$. 
\begin{equation}
  \label{eq:13}
  \pounds_{v^L_\epsilon} Q_X=0.
\end{equation}
When $\epsilon=\epsilon(t)$, these transformation are gauge invariances of the
action. Indeed, the variation of the Lagrangian density is
$\frac{d}{dt}\langle b_0,\epsilon(t) \rangle$ so that the variation of the
action vanishes for all $\epsilon(t)$ that vanish at the end points of the path
$\gamma$. In these terms, \eqref{eq:13} means that the Noether charges for the
global symmetries, including the Hamiltonian, are gauge invariant.

\section{Constrained Hamiltonian analysis of geometric actions}
\label{sec:hamilt-analys-}

Even though geometric actions are already in first order form, the Hamiltonian
analysis is not complete because of the degeneracies of the pre-symplectic two
form. In order to have explicit expressions for Poisson and Dirac brackets,
required in the context of operator quantization and extended formulations of
the theory with both group elements and coadjoint vectors as dynamical
variables, it is instructive to perform a complete constrained Hamiltonian
analysis. Conversely, such world-line actions are prime examples where Dirac's
theory comes into its own in the case of completely tractable mechanical systems
as opposed to field theories. We refer to the reviews \cite{Choquet} on Lie
groups and \cite{Hanson1976,Henneaux:1992ig} on constrained Hamiltonian systems
for more details and proofs. There is of course no claim of originality as all
results are known in one form or the other in the (mathematical) literature.

\subsection{Lie groups and algebras in local coordinates}
\label{sec:lie-groups-algebras}

We find it convenient to perform the analysis by using explicit (arbitrary)
local coordinates $g^i$ on $G$. At the same time, even though not necessary
for our purpose here, we provide in parenthesis the simplified expressions for
the objects of section \ref{sec:generalities} for the case of (subgroups) of
${\rm GL}(n)$, where the inner product $\langle \cdot,\cdot \rangle$ is the
matrix trace.

In local coordinates, the left/right translations $L_g/R_g$ are
encoded in the multiplication table $L^i(g^j,h^k)$ /
$R^i(g^j,h^k)$, while their differentials $L'_g$ / $R'_g$
needed to push-forward vector fields from $h$ to $gh/hg$ are
characterized by
\begin{equation}
  \label{eq:17}
  \frac{\partial L^i(g^l,h^k)}{\partial h^j}\quad /
  \quad\frac{\partial R^i(g^l,h^k)}{\partial h^j}. 
\end{equation}
The matrices
\begin{equation}
  \label{eq:19}
  {L^i}_j(g^l)=\left.\frac{\partial
      L^i(g^l,h^k)}{\partial h^j}\right|_{h=e}
  \quad / \quad
{R^i}_j(g^l)=\left.\frac{\partial R^i(g^l,h^k)}{\partial h^j}\right|_{h=e}, 
\end{equation}
are invertible, reduce to $\delta^i_j$ at the identity $e$, and
commute, ${L^i}_j {R^j}_k={R^i}_j {L^j}_k$.

Denoting by $e_i=\left.\frac{\partial}{\partial g^i}\right|_{e}$, the coordinate
basis for tangent vector fields at the identity, bases for the generators of
right/left translations which are the left/right invariant vector fields
$v^R_{e_i}(=ge_i)$ / $v^L_{e_i}(=e_i g)$ that reduce to $e_i$ at the identity $e$ are given by
\begin{equation}
  \label{eq:18}
  v^R_{e_i}= {L^j}_i\frac{\partial}{\partial g^j}\quad /\quad
  v^L_{e_i}= {R^j}_i\frac{\partial}{\partial g^j},
\end{equation}
Left/right invariance of these vector fields translates into
\begin{equation}
  \label{eq:61}
  \frac{\partial L^i(g,h)}{\partial g^l}{L^l}_j(h)={L^i}_{j}(gh)\quad / \quad
  \frac{\partial R^i(g,h)}{\partial g^l}{R^l}_j(h)={R^i}_{j}(hg). 
\end{equation}
These bases are mutually commuting,
\begin{equation}
  \label{eq:31}
  [v^R_{e_i},v^L_{e_j}]=0,
\end{equation}
and their rotation coefficients are determined by the Lie algebra structure
constants,
\begin{equation}
  \label{eq:20}
  [v^R_{e_i},v^R_{e_j}]=f^k_{ij}v^R_{e_k}\quad/\quad
  [v^L_{e_i},v^L_{e_j}]=-f^k_{ij}v^L_{e_k}.
\end{equation}
The left/right invariant
Maurer-Cartan forms $\theta(=g^{-1}dg),\kappa(=dg g^{-1})$ are given by
\begin{equation}
  \label{eq:21}
  \theta=e_i {(L^{-1})^i}_j dg^j\quad / \quad \kappa = e_i {(R^{-1})^i}_j dg^j,
\end{equation}
and satisfy
\begin{equation}
  \label{eq:22}
  d\theta+\frac 12 [\theta,\theta]=0\quad / \quad d\kappa -\frac 12 [\kappa,\kappa]=0,
\end{equation}
with $[e_i,e_j]=f^k_{ij}e_k$. The adjoint representation is determined
by
\begin{equation}
  \label{eq:23}
  {\rm Ad_g}e_i = e_j{(R^{-1}L)^j}_i, 
\end{equation}
with ${(R^{-1}L)^j}_i={(R^{-1})^j}_k {L^k}_i$. In the
following, we will use 
\begin{equation}
v^L_{e_i}{(R^{-1}L)^j}_l=f^j_{ik}{(R^{-1}L)^k}_l,\label{eq:44}
\end{equation}
which holds on account of \eqref{eq:31} and \eqref{eq:20}.

In order to explicitly show in local coordinates that replacing $b_{0i}$ by
$b'_{0i}=b_{0l} {(R^{-1}L)^l}_i$ amounts to replacing $g^i$ by
$g'^i=L^i(h,g)$, one uses left invariance in the form of \eqref{eq:61} and
the matrix expression for ${\rm Ad}_h{\rm Ad}_g={\rm Ad}_{hg}$.

\subsection{Legendre transform,
  primary constraints and canonical generators}
\label{sec:legendre-transf-prim}

In terms of local coordinates, the geometric action \eqref{eq:1} becomes
\begin{equation}
  \label{eq:24}
  I_G[g;b_0,X_0]=\int dt\ \big[b_{0i}{(R^{-1})^i}_j\dot g^j-b_{0i}{(R^{-1}L)^i}_j X^j_0\big]. 
\end{equation}
Denoting by $p_j$ the canonical momenta, $\{g^i,p_j\}=\delta^{i}_j$,
$\{g^i,g^j\}=0=\{p_i,p_j\}$, the primary constraints and the canonical
Hamiltonian are
\begin{equation}
  \label{eq:25}
  \tilde \phi^{b_0}_j=p_j-b_{0i}{(R^{-1})^i}_j\approx 0,\quad
  H_C=p_i{L^i}_j X^j_0\approx b_{0i}{(R^{-1} L)^i}_j X^j_0.
\end{equation}
By construction, geometric actions are then equivalent to
\begin{equation}
  \label{eq:57}
  I_G^H[g^i,p_j,\tilde u^m;b_0,X_0]=\int dt\,\big[p_j \dot q^j -H_C-\tilde u^i\tilde \phi^{b_0}_i\big]. 
\end{equation}
Models with different $b_0$'s that belong to the same coadjoint orbit are described
by equivalent actions, $I^H_G[g^i,p_j,u^m;{\rm Ad}^*_{h^{-1}}
b_0,X^0]=I^H_G[g'^i,p'_j,\tilde u'^m; b_0,X^0]$ that are related through the
canonical transformations, 
\begin{equation}
  \label{eq:58}
  g'^i =L^i(h^k,g^l),\quad p'_j={L^m}_l(g){(L^{-1})^l}_j(hg)p_m,
\end{equation}
together with
\begin{equation}
  \label{eq:99}
  \tilde u'^m=
  {L^m}_{j}(hg){(L^{-1})^j}_n (g) \tilde u^n.
\end{equation}

In the following, it turns out to be convenient not to use Darboux coordinates
$(g^i,p_j)$, but rather to change coordinates on phase space to
$g^i$ and 
\begin{equation}
\pi_j={R^k}_j p_k\label{eq:49}.
\end{equation}
The fundamental Poisson brackets in terms
of these coordinates are
\begin{equation}
  \label{eq:26}
  \{g^i,g^j\}=0,\quad\{g^i,\pi_j\}={R^i}_j,\quad \{\pi_i,\pi_j\}=f^k_{ij}\pi_k.
\end{equation}
In particular, $\{\pi_i,\cdot\}=\pi_k
f^k_{ij}\frac{\partial}{\partial\pi_j}-v^{L}_{e_i}$.
Under the canonical transformation designed to compensate a change of the orbit
representative $b_0$, these variables transform as
\begin{equation}
  \label{eq:65}
  \pi'_i=\pi_l {(R^{-1}L)^l}_i(h).
\end{equation}
In the
mathematical literature, when considering $\pi_i$ as coordinates on $\mathfrak
g^*$, $\pi=\pi_i e^{*i}$, the above Poisson brackets for the $\pi_i$'s are
referred to as the Lie-Poisson bracket or Kirillov-Kostant-Souriau bracket on
$\mathfrak g^*$.

The primary constraints are equivalent to
\begin{equation}
  \label{eq:27}
  \phi^{b_0}_i=\pi_i-b_{0i}\approx 0,  
\end{equation}
while the canonical Hamiltonian may be chosen as
\begin{equation}
  \label{eq:28}
  H^\pi_{X_0}=\pi_i {(R^{-1}L)^j}_i X^j_0.
\end{equation}
By construction, the theory defined by the geometric action $I[g;b_0,X^0]$ is
equivalent to the one defined by
\begin{equation}
  \label{eq:30}
  I^H_G[g,\pi,u;b_0,X^0]=\int dt\ \big[\pi_i
    {(R^{-1})^i}_j
    \dot g^j-H^\pi_{X_0}-u^i\phi^{b_0}_i\big],
\end{equation}
where $u^i$ are Lagrange multipliers that may be considered as elements of
$\mathfrak g$, $u=u^i e_i$, that transform in the adjoint representation,
\begin{equation}
  \label{eq:100}
  u'^i={(R^{-1}L)^i}_j(h)u^j. 
\end{equation}
When using the second of \eqref{eq:22},
variations with respect to the dynamical variables gives,
\begin{multline}
  \label{eq:95}
  \delta I_G^H=\int dt\, \Big[\delta\pi_i[{(R^{-1})^i}_j \dot
  g^j-{(R^{-1}L)^i}_j X_0^j-u^i]-\delta u^i\phi^{b_0}\\
  +[-\dot
  \pi_i{(R^{-1})^i}_j +\pi_i f^i_{kl}{(R^{-1})^k}_j{(R^{-1})^l}_m\dot g^m-\pi_i\partial_j{(R^{-1}L)^i}_m X^m_0 ]\delta g^j
  \Big]. 
\end{multline}
The Euler-Lagrange equations with respect to $g^j$ may then be simplified using
\eqref{eq:44}, and the associated dynamics is
provided by the primary constraints \eqref{eq:27}, together with the
Hamiltonian evolution equations
\begin{equation}
  \label{eq:29}
  \begin{split}
    \dot g^i&=\{g^i,H^\pi_{X_0}+u^j\phi^{b_0}_j\}={L^i}_j X^j_0+{R^i}_j u^j,\\
    \dot \pi_i&=\{\pi_i,H^\pi_{X_0}+u^j\phi^{b_0}_j\}=\pi_k f^k_{ij}u^j.
  \end{split}
\end{equation}

In the Hamiltonian formalism, the Noether charges $Q^\pi_X$ 
\begin{equation}
  \label{eq:32}
  Q^\pi_{X}=\pi_i{(R^{-1}L)^i}_j X^j,
\end{equation}
canonically generate right translations and do not act on the $\pi_i$,
\begin{equation}
  \label{eq:50}
  \delta_X g^i=v^{R}_X (g^i)={L^i}_j
  X^j=\{g^i,Q^\pi_X\},\quad \delta_X \pi_i=\{\pi_i,Q^\pi_X\}=0.
\end{equation}
They form a Poisson bracket realization of $\mathfrak g$,
\begin{equation}
  \label{eq:33}
  \{Q^\pi_{X_1},Q^\pi_{X_2}\}=Q^\pi_{[X_1,X_2]}. 
\end{equation}

\subsection{Dirac algorithm: first and second class constraints}
\label{sec:dirac-algorithm}

The preservation in time of the primary constraints,
$\{\phi_i,H^\pi_{X_0}+u^j\phi^{b_0}_j\}\approx 0$ leads to
\begin{equation}
  \label{eq:34}
  b_{0k} f^k_{ij} u^j=0.
\end{equation}
To solve this equation, one considers vectors 
${e_a}^i$ that constitute a complete set of zero eigenvectors of
the matrix $C_{ij}=b_{0k} f^k_{ij}$,
\begin{equation}
  \label{eq:36}
  b_{0k} f^k_{ij} v^j=0 \iff v^j=v^a{e_a}^j,
\end{equation}
and introduces an associated change of basis in
$\mathfrak g$ and $\mathfrak g^*$ defined through constant
matrices ${e_a}^i,{e_A}^i$,
${e^a}_i,{e^A}_i$ such that 
\begin{equation}
  \label{eq:35}
  {e_a}^i {e^b}_i=\delta_a^b,\quad{e_A}^i {e^b}_i=0,\quad
  {e_A}^i {e^B}_i=\delta^B_A,\quad{e_a}^i{e^a}_j+{e_A}^i{e^A}_j=\delta^i_j.
\end{equation}
In the following, we will use these constant vielbeins and their
inverse to transform quantities with greek indices into the same
quantities with small and capital Latin indices. In terms of the new
basis, the little algebra $\mathfrak{h}_0$ is determined by vectors
such that $v^A=0$. Furthermore, 
\begin{equation}
f^C_{ab}=0,\quad C_{ab}= 0,\quad C_{aB}=0\label{eq:51}
\end{equation}
while the matrix 
\begin{equation}
C_{AB}=b_{0c}f^c_{AB}+b_{0C}f^C_{AB}\label{eq:52}
\end{equation}
is invertible, with inverse denoted by $(C^{-1})^{AB}C_{BC}=\delta^A_B$. Equation
\eqref{eq:34} leaves the Lagrange multipliers $u^a$ undetermined and
sets to zero the remaining Lagrange multipliers $u^A=0$. There are thus no
secondary constraints, while the primary constraints split into 
first and second class constraints
\begin{equation}
  \label{eq:37}
  \phi^{b_0}_a=\pi_a-b_{0a}\approx 0,\quad
  \phi^{b_0}_A=\pi_A-b_{0A}\approx 0,
\end{equation}
with
\begin{equation}
  \label{eq:39}
  \begin{split}
  & \{\phi^{b_0}_a,\phi^{b_0}_b\}=f^c_{ab}\phi^{b_0}_c,\quad
  \{\phi^{b_0}_a,\phi^{b_0}_B\}=f^c_{aB}\phi^{b_0}_c+f^C_{aB}\phi^{b_0}_C,\\
  & \{\phi^{b_0}_A,\phi^{b_0}_B\}=f^c_{AB}\phi^{b_0}_c+f^C_{AB}\phi^{b_0}_C+C_{AB}.
\end{split}
\end{equation}
The gauge transformations are generated by the first class
constraints,
\begin{equation}
  \label{eq:55}
  \delta_{\epsilon} g^i=\{g^i,\phi^{b_0}_a
  \epsilon^a\}={R^i}_a\epsilon^a,\quad \delta_\epsilon \pi_i=\{\pi_i,\phi^{b_0}_a
  \epsilon^a\}\approx 0,
\end{equation}
with $\epsilon^a=\epsilon^a(t)$. The Hamiltonian $H^\pi_{X_0} $ is
first class while the total Hamiltonian
\begin{equation}
  \label{eq:40}
 H^T_{X_0}=H^\pi_{X_0}+u^a\phi_a,
\end{equation}
is also the extended Hamiltonian since there are no secondary
constraints. In terms adapted to the classification of the
constraints, one has 
\begin{equation}
  \begin{split}
  \label{eq:41}
  I^H_G[g^i,\pi_b,\pi_B,u^c,u^C;b_0,X^0]&=\int dt\ \big[a^H_i\dot g^i
  -H^\pi_{X_0}-u^a\phi_a-u^A\phi_A\big],\\
    a^H_i&=\pi_a{(R^{-1})^a}_i+\pi_A{(R^{-1})^A}_i.
  \end{split}
\end{equation}

\subsection{Reduced theory and Dirac brackets}
\label{sec:reduced-theory-dirac}

At this stage, one may solve the second class constraints in the action and
eliminate the $\pi_A$ in favor of $b_{0A}$. The reduced theory becomes
\begin{equation}
  \label{eq:42}
  \begin{split}
  I^R_G[g^i,\pi_a,u^a;b_0,X_0]&=\int dt\ \big[a^R_i \dot g^i 
    -H^R_{X_0}-u^a\phi^{b_0}_a\big],\\
    a^R_i&=\pi_a{(R^{-1})^a}_i +b_{0A}{(R^{-1})^A}_i,\\
    H^R_{X_0}&=\pi_a{(R^{-1}L)^a}_j X_0^j+b_{0A}{(R^{-1}L)^A}_j X_0^j,
  \end{split}
\end{equation}
with associated reduced brackets defined by the
inverse of the symplectic form $\Omega^R=da^R$, with $a_R=a^R_i d g^i$. More
explicitly, 
\begin{equation}
  \label{eq:45}
  \Omega^R=d\pi_a{(R^{-1})^a}_i dg^i+\frac 12 C_{AB}{(R^{-1})^A}_i {(R^{-1})^B}_j dg^i dg^j.
\end{equation}
and 
\begin{equation}
  \label{eq:101}
  \{g^i,g^j\}^R={R^i}_A (C^{-1})^{AB}{R^j}_B,\quad \{g^i,\pi_a\}^R={R^i}_a,\quad \{\pi_a,\pi_b\}^R=0.
\end{equation}
When using \eqref{eq:44}, it follows that the dynamics of the reduced
theory is
\begin{equation}
  \label{eq:46}
  \begin{split}
    \phi^{b_0}_a& = 0,\\
    \dot g^i&=
                \{g^i,H^R_{X_0}\}^R+u^b\{g^i,\phi^{b_0}_b\}^R\approx 
                {L^i}_j X^j_0+{R^i}_b u^b,\\
    \dot \pi_a&= \{\pi_a,H^R_{X_0}\}^R+u^b\{\pi_a,\phi^{b_0}_b\}^R\approx 0.
  \end{split}
\end{equation}
Note that if one also eliminates the first class constraints by solving them in the
action, one recovers $I_G[g;b_0,X_0]$ in \eqref{eq:1} with degenerate two-form
$\Omega$.

Alternatively, one may work with Dirac brackets and keep the variables
$\pi_A,u^A$ together with the second-class constraints $\phi^{b_0}_A\approx 0$
for reasons of Lie algebra covariance. In this case, the Dirac brackets are
given by
\begin{equation}
  \label{eq:43}
  \begin{split}
    & \{g^i,g^j\}^*=  {R^i}_A (C^{-1})^{AB}{R^j}_B,\\
    & \{g^i,\pi_a\}^*= {R^i}_a-{R^i}_A (C^{-1})^{AB}(f^c_{Ba}\phi^{b_0}_c+f^C_{Ba}\phi^{b_0}_C),\\
    & \{\pi_a,\pi_b\}^*=f^c_{ab}\phi^{b_0}_c
      -(f^c_{aA}\phi^{b_0}_c+f^C_{aA}\phi^{b_0}_C)(C^{-1})^{AB}(f^d_{Bb}\phi^{b_0}_d+f^D_{Bb}\phi^{b_0}_D),
  \end{split}
\end{equation}
which agree with the reduced brackets on the constraint surface,
$\{\cdot,\cdot\}^*\approx \{\cdot,\cdot\}^R$, while the additional Dirac brackets all vanish, 
\begin{equation}
  \label{eq:47}
  \{g^i,\pi_A\}^*= 0,\quad \{\pi_a,\pi_B\}^*= 0,\quad \{\pi_A,\pi_B\}^*= 0. 
\end{equation}

A point on the coadjoint orbit of a given covector $b_0$ can be parametrized by
the group element needed to reach it, 
\begin{equation}
  \label{eq:68}
  b^g_0= {\rm Ad}^*_{g^{-1}} b_0.
\end{equation}
The time dependence of such a covector is given by
\begin{equation}
  \label{eq:63}
  \frac{db^g_0}{dt}=-{\rm Ad}^*_{g^{-1}}({\rm ad}^*_{i_V\kappa}b_0).
\end{equation}
When using the equations of motion \eqref{eq:46} together with the fact that
$u^b e_b$ belongs to the little algebra of $b_0$, it follows that
\begin{equation}
  \label{eq:66}
  \frac{db^g_0}{dt}=-{\rm ad}^*_{X_0} b. 
\end{equation}

\subsection{Unconstrained model}
\label{sec:extended-model}

One may decide to use the Lie algebra covariant form
\eqref{eq:30} of the model,
\begin{equation}
  \label{eq:53}
  I^H_G[g,\pi,u;b_0,X_0]=\int_\gamma
  \big[\langle \pi,\kappa \rangle
  -dt (H^\pi_{X_0}+\langle \phi^{b_0},u \rangle)\big],
\end{equation}
where
\begin{equation}
  \label{eq:54}
  H^\pi_{X_0}=\langle {\rm Ad}^*_g\pi,X_0 \rangle,\quad
    \phi^{b_0}=\pi-b_0,
\end{equation}
without explicitly splitting into first and second class constraints.
As seen above, the latter can always be done once $b_0$ is fixed and the
little algebra has been worked out.

One may also go a step further and drop the constraints $\phi^{b_0}$ to study
the unconstrained model
\begin{equation}
  \label{eq:38}
  I^U_G[g,\pi;X_0]=\int_\gamma \big[\langle \pi,\kappa \rangle-dt H^\pi_{X_0}\big],
\end{equation}
with Poisson brackets given in \eqref{eq:26} and Hamiltonian evolution equations
that simplify to
\begin{equation}
  \label{eq:60}
  \dot g^i=\{g^i,H^\pi_{X_0}\}={L^i}_j X^j_0\iff
  i_V \kappa={\rm Ad}_g X_0,\quad \dot \pi_i=\{\pi_i,H^\pi_{X_0}\}=0. 
\end{equation}
It follows that, besides the $Q^\pi_{X}$, the $\pi$'s themselves are constants of the
motion, 
\begin{equation}
  \label{eq:64}
  \pi_i=b_{0i},
\end{equation}
with $b_{0i}$ constant. On these level sets, one can study Hamiltonian
reduction. This amounts to performing the analysis in the previous section.

\section{Geometric action for extended BMS$_4$ group} 
\label{sec:geom-acti-bms_4}

\subsection{Group and algebra}
\label{sec:group}

The extended BMS$_4$ group is a semi-direct product group of the form
\begin{equation}
  \label{eq:62}
  \mathcal{S}_\sigma=G\ltimes_{\sigma} A,\quad 
\end{equation}
with $A$ and abelian ideal,
\begin{equation}
  \label{eq:69}
  (g_1,\alpha_1)\cdot (g_2,\alpha_2)=(g_1\cdot g_2, \alpha_1+\sigma_{g_1}\alpha_2). 
\end{equation}
The non-abelian factor $G$ corresponds to conformal coordinate
transformations on the complex plane minus the origin, $g=(f,\bar f)$,
\begin{equation}
  \label{eq:89}
  z'(z)=f(z),\quad
  \bar z'(\bar z)=\bar f(\bar z),
\end{equation}
with group law on the level of $f,\bar f$ defined by composition.

The abelian ideal $A$ consists of real fields $T$ of conformal dimensions
$(-\frac 12,-\frac 12)$, with
\begin{equation}
  \label{eq:92}
  (\sigma_g T)(z',\bar z')=\big(\frac{\partial z}{\partial z'}\big)^{-\frac 12}
  \big(\frac{\partial \bar z}{\partial \bar z'}\big)^{-\frac 12}T(z,\bar z). 
\end{equation}
More generally $G$ acts on conformal fields of dimensions $(h,\bar h)$ as
\begin{equation}
  \label{eq:93}
  (\sigma_g\phi^{h,\bar h})(z',\bar z')=\big(\frac{\partial z}{\partial z'}\big)^{h}
  \big(\frac{\partial \bar z}{\partial \bar z'}\big)^{\bar h}\phi^{h,\bar h}(z,\bar z). 
\end{equation}
while $\Sigma_X$ acts as
\begin{equation}
  \label{eq:72}
  (Y,\bar Y)\cdot \phi^{h,\bar h}= -\big[Y\partial
  +\bar Y\bar\partial+h \partial Y
  +\bar h\bar\partial\bar Y \big]\phi^{h,\bar h}.
\end{equation}

The associated Lie algebra $\mathfrak{bms}^E_4$ is of the
form $\mathfrak{g}\oright_\Sigma A$,
\begin{equation}
  \label{eq:70}
  [(X_1,\alpha_1),(X_2,\alpha_2)]=([X,Y],\Sigma_{X_1}\alpha_2-\Sigma_{X_2}\alpha_1), 
\end{equation}
where $\Sigma_X$ is the differential of $\sigma_g$ and we identify the Lie algebra
elements of $A$ with elements of $A$ itself. The Lie algebra of $G$ is described
by chiral fields $Y,\bar Y$ of conformal dimensions $(-1,0)$ and $(0,-1)$,
\begin{equation}
  \label{eq:91}
  \bar\partial Y=0=\partial\bar Y,
\end{equation}
while the $\mathfrak{bms}^E_4$ Lie bracket is explicitly given by 
\begin{equation}
  \begin{split}
  \label{eq:90}
    & [(Y_1,\bar Y_1,T_1),(Y_2,\bar Y_2,T_2)]=-\big(\hat Y,\hat{\bar Y},\hat T),\\
    & \hat Y=Y_1\partial
  Y_2-(1\leftrightarrow 2),\quad\hat{\bar Y}=
      \bar Y_1\bar\partial \bar Y_2-(1\leftrightarrow 2),\\
    & \hat T=Y_1\partial T_2-\frac 12 \partial Y_1T_2+{\rm c.c.}-(1\leftrightarrow 2).
  \end{split}
\end{equation}

\subsection{Adjoint and coadjoint representation}
\label{sec:adjo-coadj-repr}

The adjoint action is of the form
\begin{equation}
{\rm Ad}_{(g,\alpha)} (X, \beta) = ({\rm Ad}_g X, \sigma_g \beta -
    \Sigma_{{\rm Ad}_g X} \alpha), \label{Adjoint}
\end{equation}
where ${\rm Ad}_g X$ is given by $\big(g\cdot
(Y,\bar Y)\big)
(x')= \big({Y}'(x'),\bar{Y}(x')\big)$ with 
\begin{equation}
  \label{eq77} 
  Y'(z')=\big(\frac{\partial z}{\partial z'}\big)^{-1}Y(z),\quad
  {\bar Y}'(\bar z')= \big(\frac{\partial \bar z}{\partial \bar z'}\big)^{-1}{\bar Y}(z). 
 \end{equation}
If $\alpha=\mathcal{T}_1$, $\beta=\mathcal{T}_2$,
$\sigma_g\beta-\Sigma_{{\rm Ad}_g X}\alpha$ is given by
\begin{equation}
  \label{eq:76}
  T'(x')=\big(\frac{\partial z}{\partial z'}\big)^{-\frac{1}{2}}
  \big(\frac{\partial \bar z}{\partial \bar z'}\big)^{-\frac{1}{2}}\Big(
  T_2+\big({{Y}}\partial T_1- \frac{1}{2}
  T_1 \partial {{Y}} + {\rm c.c.}\big)\Big)(x).
\end{equation}

The dual space to the Lie algebra is of the form $\mathfrak{g}^*\oplus
A^*$, with non-degenerate pairing denoted by
\begin{equation}
\langle (j, p), (X, \alpha)\rangle = \langle j, X \rangle + \langle p,
\alpha \rangle.
\label{abstract pairing}
\end{equation}
In terms of 
\begin{equation}
  \label{eq:102}
  \times: A\oplus A^*\to \mathfrak g^*,\quad \langle \alpha \times p, X \rangle = \langle p, \Sigma_X \alpha
  \rangle,
\end{equation}
and $\sigma^*$, the dual realization associated with $\sigma$,
$\sigma^*: G\times A^*\to A^*$,
$\langle \sigma^*_g p, \alpha \rangle = \langle p, \sigma_{g^{-1}} \alpha\rangle$,
the coadjoint actions of the group and algebra are of the form
\begin{equation}
  \begin{split}
    {\rm Ad}^*_{(g, \alpha)} (j, p) &= ({\rm Ad}^*_g j + \alpha \times
\sigma^*_g p, \sigma^*_g p),\\
    \label{CoAdjointe}
{\rm ad}^*_{(X, \alpha)} (j, p) &= ({\rm ad}^*_X j + \alpha \times p,
\Sigma^*_X p).
  \end{split}
\end{equation}
In the case of $(\mathfrak{bms}^E_4)^*$, elements are denoted by
$([J],[\bar{J}],P)$. Here $J,\bar{J}$ have conformal dimensions $(1,2)$ and
$(2,1)$, while $P$ has dimensions $(\frac 32,\frac 32)$, with pairing given by
\begin{equation}
  \langle([J],[\bar{J}], P), (\mathcal Y, \bar{\mathcal Y}, \mathcal T)
  \rangle = \int di(x)\, [\bar{J} Y+
  J\bar{Y}+ P T]. \label{eq:312}
\end{equation}
As discussed in more details in
\cite{Barnich:2021dta}, if we assume that conformal fields may be expanded in
terms of suitable series in $z,\bar z$,
\begin{equation}
  \label{eq:103}
  \phi_{h,\bar h}(z,\bar z)=\sum_{k,l}a_{k,l}z^{-h-k}\bar z^{-\bar h-l},
\end{equation}
the integral correponds to taking
residues in $z,\bar z$: $\int di(x)=\int dz d\bar z$ with 
\begin{equation}
  \label{eq:109}
  \begin{split}
    & \int dz\, \phi_{h,\bar h}(z,\bar z)={\rm Res}_z\phi_{h,\bar h}(z,\bar z)=\sum_l a_{1-h,l}\bar z^{-\bar h-l},\\
    & \int d\bar z\, \phi_{h,\bar h}(z,\bar z)={\rm Res}_{\bar z}\phi_{h,\bar h}(z,\bar z)=\sum_k a_{k,1-\bar h} z^{-\bar h-k}.
  \end{split}
\end{equation}
Because $Y,\bar Y$ are chiral fields, one has to consider
equivalence classes, $J\sim J+\partial L$,
$\bar{J}\sim \bar{J}+\bar \partial \bar L$ with $L,\bar L$ of dimensions 
$(0,2)$ and $(2,0)$. In these terms,
the coadjoint representation is given by 
\begin{equation}
  \label{eq:65c}
  \begin{split}
    &J'(x')=\big(\frac{\partial z}{\partial z'}\big)^{1}
    \big(\frac{\partial \bar z}{\partial \bar z'}\big)^{2}\Big(J-
    (\frac{1}{2}{{T}} \bar\partial {{P}} +\frac{3}{2}
    \bar\partial
    {{T}} {{P}}  )\Big)(x),\\
    &{\bar J}'(x')=\big(\frac{\partial z}{\partial z'}\big)^{2}
    \big(\frac{\partial \bar z}{\partial \bar z'}\big)^{1}\Big({\bar J}-
    (\frac{1}{2} {{T}} \partial {{P}} +
    \frac{3}{2}\partial
    {{T}} {{P}}  )\Big)(x),\\
    &P'(x')=\big(\frac{\partial z}{\partial z'}\big)^{\frac{3}{2}}
    \big(\frac{\partial \bar z}{\partial \bar
      z'}\big)^{\frac{3}{2}}P(x).
  \end{split}
\end{equation}
Representatives for the equivalence classes
$[\bar J],[J]$ are given
by
\begin{equation}
  \label{eq:104}
  \bar J(z,\bar z)=\bar J(z)\delta(\bar z,0),\quad J(z,\bar z)=\delta(z,0)J(\bar z), 
\end{equation}
where $\bar J(z)=\int d\bar z\, \bar J(z,\bar z)$ is of conformal dimension $h=2$,
while $J(\bar z)=\int dz\, J(z,\bar z)$ is of
conformal dimension $\bar h=2$. This follows from the fact that the only
term in a series in $z,\bar z$ that cannot be written as a
$\bar \partial/\partial$
derivative is $\bar z^{-1}/ z^{-1}$ and from the series expansion of the delta
function:   if
\begin{equation}
  \label{eq:105}
  \bar J(z,\bar z)=\sum_{k,l}\bar J_{k,l}z^{-2-k}\bar z^{-1-l},\quad J(z,\bar z)=
  \sum_{k,l} J_{k,l}z^{-1-k}\bar z^{-2-l},
\end{equation}
we have
\begin{equation}
  \label{eq:106}
  \bar J(z,\bar z)=\sum_k\bar J_{k,0}z^{-2-k}\bar z^{-1}+\bar\partial\bar L,\quad
  J(z,\bar z)=\sum_l J_{0,l}z^{-1}\bar z^{-2-l}+\partial L,
\end{equation}
with
\begin{equation}
  \label{eq:107}
  \bar L=-\sum_{k,l\neq 0}\bar J_{k,l} l^{-1}z^{-2-k}\bar z^{-l},\quad
  L=-\sum_{k\neq 0,l} J_{k,l} k^{-1} z^{-k}\bar z^{-2-l}.
\end{equation}
and furthermore, 
\begin{equation}
  \label{eq:108}
  \delta(z,w)=\sum_{k}z^{k-1}w^{-k},\quad \delta(\bar z,\bar w)=\sum_{k}\bar z^{k-1}\bar w^{-k}. 
\end{equation}

\subsection{Unconstrained model for extended BMS$_4$}
\label{sec:unconstrained-model}

For a semi-direct product group of the form \eqref{eq:62}, the left/right invariant
Maurer-Cartan forms $\theta_{g,\alpha}/\kappa_{g,\alpha}$ are given by
\begin{equation}
  \label{eq:77}
  \theta_{g,\alpha}=(\theta_g,\sigma_{g^{-1}}d\alpha)\quad /\quad\kappa_{g,\alpha}=(\kappa_g,d\alpha-\Sigma_{\kappa_g}\alpha),
\end{equation}
where $\theta_g/\kappa_g$ denote the left/right invariant Maurer-Cartan forms of
the non-abelian group $G$. It then follows from \eqref{eq:102} that the kinetic
term of the unconstrained model \eqref{eq:38} is 
\begin{equation}
  \label{eq:78}
  \langle \pi,\kappa \rangle=\langle j-\alpha\times p,\kappa_g \rangle+\langle  p,d\alpha \rangle,
\end{equation}
while the Hamiltonian $\langle  \pi,{\rm Ad}_gX_0 \rangle$ is
\begin{equation}
  \label{eq:79}
  H_{(X_0,\beta_0)}=\langle j-\alpha\times p,{\rm Ad}_gX_0 \rangle+\langle p,\sigma_g\beta_0\rangle.
\end{equation}
Furthermore, the identity
$\langle \pi,\kappa \rangle=\langle {\rm Ad}^*_{g^{-1}}\pi ,\theta \rangle$ becomes
\begin{multline}
  \label{eq:110}
  \langle j-\alpha\times p,\kappa_g \rangle+\langle p,d\alpha \rangle=\langle {\rm Ad}^{*}_{(g^{-1},-\sigma_{g^{-1}}\alpha)}(j,p),(\theta_g,\sigma_{g^{-1}}d\alpha) \rangle\\
  =\langle {\rm Ad}^*_{g^{-1}}j-\sigma_{g^{-1}}\alpha\times \sigma^*_{g^{-1}}p,\theta_g \rangle+\langle \sigma^*_{g^{-1}}p,\sigma_{g^{-1}}d\alpha \rangle,
\end{multline}
while the Hamiltonian may also be written as
$\langle {\rm Ad}^*_{g^{-1}} j, X_0\rangle$.

For conformal coordinate transformations, $z\mapsto z'=f(z)$,
$\bar z\mapsto z'=\bar f(\bar z)$ the left/right invariant Maurer-Cartan forms
are
\begin{equation}
  \label{eq:81}
  \theta_g=(\frac{1}{f'}df\frac{\partial}{\partial z},{\rm c.c.})\quad/\quad
  \kappa_g=(df\circ f^{-1}\frac{\partial}{\partial z},{\rm c.c.}).
\end{equation}
As a consequence, the unconstrained model \eqref{eq:38} for the extended BMS$_4$
group may be written either as
\begin{multline}
  \label{eq:114}
  I^U_{{\rm BMS}^E_4}[f,\bar f,T,J,\bar J,P; Y_0,\bar{Y}_0,T_0]=\\\int dt dzd\bar z\,\Big(
  \big[\bar{J}+(\frac{1}{2}{{T}}\partial {{P}} +
  \frac{3}{2}\partial
  {{T}}{{P}} )\big]
  \big(\big[\dot f-(f'Y_0)\big]\circ f^{-1}\big)+{\rm c.c.}\\+P\dot T-
  P\big(\big[(f'\bar f')^{\frac 12}{T}_0\big]\circ(f^{-1},\bar
  f^{-1})\big)\Big),
\end{multline}
or, in a chiral boson like form, as 
\begin{multline}
  \label{eq:85}
  I^U_{{\rm BMS}^E_4}[f,\bar f,T,J,\bar J,P; Y_0,\bar{Y}_0,T_0]=\\\int dt dzd\bar z\,\Big(
  \big[\big(\bar J+(\frac 12 T\partial P+\frac 32 \partial TP)\big)\circ (f,\bar
  f)\big] \big[f'\bar f'
  \dot f-(f')^2\bar f' Y_0\big]+{\rm c.c.}\\+
  P\dot T-[P\circ(f,\bar f)](f'\bar f')^{\frac{3}{2}}T_0\Big).
\end{multline}
That these forms of the model are equivalent may also be shown by replacing in
the integral in \eqref{eq:114} of the relevant terms the dummy variables
$z,\bar z$ by $z',\bar z'$ and then performing the change of coordinates
$z'=f(z),\bar z'=\bar f(\bar z)$.

By construction, the model is invariant under infinitesimal right BMS$_4$
transformations,
\begin{equation}
  \label{eq:87}
  \delta_R f=f'Y_R,\quad \delta_R\bar f=\bar f'\bar{Y}_R,\quad
  \delta_R T=\big[(f'\bar f')^{\frac 12}T_R\big]\circ(f^{-1},\bar f^{-1}),
\end{equation}
provided that
\begin{equation}
  \label{eq:88}
  \begin{split}
  &\dot Y_R= Y_0\partial Y_R-Y_R\partial Y_0,\quad
  \dot{\bar{Y}}_R=
  \bar{Y}_0\bar\partial\bar{Y}_R-\bar{Y}_R\bar\partial \bar{Y}_0,\\
  &\dot T_R=Y_0\partial T_R-\frac 12 \partial Y_0 T_R-Y_R\partial T_0+\frac 12 \partial Y_R T_ 0+{\rm c.c.}.
  \end{split}
\end{equation}
The direct check of invariance on the form \eqref{eq:114} uses
$\delta_R f^{-1}=-Y_R\circ f^{-1}$ and the associated complex conjugate relation, as
well as spatial integrations by parts.

The associated equations of motion are explicitly given by
\begin{equation}
  \label{eq:59}
  \begin{split}
  & \dot f= f'Y_0,\quad \dot{\bar f}= \bar f'\bar Y_0,\quad 
    \dot T= \big[(f'\bar f')^{\frac 12}{T}_0\big]\circ(f^{-1},\bar f^{-1}),\\
  & \dot P=0,\quad \dot{\bar J}(z)=0,\quad\dot J(\bar z) = 0. 
  \end{split}
\end{equation}
while the Poisson brackets $\{\pi_i,\pi_j\}=f^{k}_{ij}\pi_k$ read explicitly
\begin{equation}
  \label{eq:94}
  \begin{split}
    \{\bar J(z),P(w,\bar w)\}&=[\frac 32 \partial_w\delta(z,w)
                                      + \delta(z,w)\partial_w]P(w,\bar w),\\
    \{J(z),P(w,\bar w)\}&=[\frac 32 \partial_{\bar w}\delta(\bar z,\bar w)+
                          \delta(\bar z,\bar w)\partial_{\bar w}]P(w,\bar w),\\
    \{\bar J(z),\bar J(w)\}&=[2\partial_w\delta(z,w)
                             +\delta(z,w)\partial_w]\bar J(w),\\
    \{J(\bar z),J(\bar w)\}&=[2\partial_{\bar w}\delta(\bar z,\bar w)+
                             \delta(\bar z,\bar w)\partial_{\bar w}]J(\bar w),\\
    \{\bar J(z),J(w)\}&=0,\\
    \{P(z,\bar z),P(w,\bar w)\}&=0.
      \end{split}
\end{equation}
This can be shown from
$\{\pi_i X_1^i,\pi_j X^j_2\}=\pi_k [X_1,X_2]^k$. An manifestly
skew-symmetric form of the brackets may be obtained by using the relations
\begin{equation}
  \label{eq:98}
  \partial_z\delta(z,w)=-\partial_w\delta(z,w),\quad F(w)\partial_z\delta(z,w)=F(z)\partial_z\delta(z,w)+\partial_zF(z)\delta(z,w).
\end{equation}

\section{Relation to asymptotically flat gravity at null infinity and celestial
  holography}
\label{sec:relat-asympt-flat}

The Lie-Poisson or Kirillov-Kostant-Souriau brackets \eqref{eq:94} are the
classical analogs of the operator product expansions that have recently appeared
in the context of celestial holography
\cite{Kapec2017,Barnich:2017ubf,Fotopoulos2020,Donnay2021}.

In the spirit of effective field theories, since the symmetry group of
asymptotically flat spacetimes at null infinity is the BMS$_4$ group, the
current algebra and conserved charges of the model are expected to reproduce the
behavior of the currents and charges of these spacetimes
\cite{Wald:1999wa,Barnich:2011mi,Barnich:2013axa}. In particular, when choosing
the Hamiltonian associated to (retarded) time-translations $t=u$, $X_0=(0,0,1)$,
$Y_0=0=\bar Y_0$, $T_0=1$,
\begin{equation}
  \label{eq:86}
  H_{(0,0,1)}=\int dudzd\bar z\, {P}\big[(f'\bar f')^{\frac 12}\circ(f^{-1},\bar f^{-1})\big]
  =\int dudzd\bar z\, [P\circ(f,\bar f)](f'\bar f')^{\frac{3}{2}},
\end{equation}
the time dependence of the symmetry generators of right translations is
determined by
\begin{equation}
  \label{eq:113}
  \dot Y_R=0=\dot{\bar{Y}}_R,\quad
  \dot T_R=\frac 12 (\partial Y_R+\bar\partial \bar Y_R).
\end{equation}
This is consistent with the time dependence of the leading parts of the
asymptotic symmetries generators in the gravitational computation at future null
infinity. The equations for the group elements in \eqref{eq:59} simplify to
\begin{equation}
  \label{eq:96}
  \dot f=0=\dot{\bar f},\quad \dot T
  =[f'\bar f']^{\frac 12}\circ(f^{-1},\bar f^{-1}),
\end{equation}
so that $T=T(z,\bar z,0)+u[f'\bar f']^{\frac 12}\circ(f^{-1},\bar f^{-1})$, while
those for the coadjoint vectors are unchanged. Furthermore, in the constraint
model with points on the coadjoint orbits described by
$b^g_0={\rm Ad}^*_{g^{-1}}b_0$,
\begin{equation}
  \begin{split}
    \label{eq:67}
    & J_0^g=\big[J_0+(\frac 12 T\bar \partial P_0+\frac 32 \bar \partial TP_0)\big]\circ (f,\bar
    f) (\bar f')^2 f',\\
    & \bar J_0^g=\big[\bar J_0+(\frac 12 T\partial P_0+\frac 32 \partial TP_0)\big]\circ (f,\bar
    f) (f')^2\bar f',\\
    & P^g_0=P_0\circ(f,\bar f)](f'\bar f')^{\frac{3}{2}},
  \end{split}
\end{equation}
it follows from \eqref{eq:66} or by direct computation that
\begin{equation}
  \label{eq:71}
  \frac{dP^g_0}{du}=0,\quad \frac{dJ^g_0}{du}=\frac 12 \bar{\partial} P^g_0,\quad
  \frac{d\bar J^ g_0}{du}= \frac 12 \partial P^g_0.
\end{equation}

If $\sigma^0$, $\Psi^0_4$, $\Psi^0_3$, $\Psi^0_2$, $\Psi^0_1$ are the
leading components of shear and of suitable components of the Weyl tensor in the
Newman-Penrose description of asymptotically flat spacetimes at null infinity
and $\mathrm{f}=T+\frac 12 u(\partial Y+\bar\partial Y)$, the former transform under $\mathfrak{bms}^E_4$ as
\begin{equation}
  \label{eq:73}
  \begin{split}
    -\delta\sigma^0&=(\mathrm{f}\partial_u+Y\partial+\bar Y\partial +\frac 32 \bar\partial \bar Y-\frac 12\partial Y)\sigma^0-\bar\partial^2 \mathrm{f},\\
    -\delta\Psi^0_4&=(\mathrm{f}\partial_u+Y\partial+\bar Y\partial +\frac 52 \partial Y+\frac 12 \bar\partial \bar Y)\Psi^0_4,\\
    -\delta\Psi^0_3&=(\mathrm{f}\partial_u+Y\partial+\bar Y\partial +2 \partial Y+\bar\partial \bar Y)\Psi^0_3+\Psi^0_4\bar\partial \mathrm{f},\\
    -\delta\Psi^0_2&=(\mathrm{f}\partial_u+Y\partial+\bar Y\partial +\frac 32 \partial Y+\frac 32 \bar\partial \bar Y)\Psi^0_2+2\Psi^0_3\bar\partial \mathrm{f},\\
    -\delta\Psi^0_1&=(\mathrm{f}\partial_u+Y\partial+\bar Y\partial +\partial Y+2\bar\partial \bar Y)\Psi^0_1+3\Psi^0_2\bar\partial \mathrm{f}.
  \end{split}
\end{equation}
Furthermore, they are related
through
\begin{equation}
  \label{eq:74}
  \Psi^0_4=-\partial_u^2\bar\sigma^0,\quad\Psi^0_3=-\bar \partial \partial_u\bar\sigma^0,\quad \Psi^0_2-\bar\Psi^0_2=\partial^2\sigma^0-\bar\partial^2\bar\sigma^0+\bar\sigma^0\partial_u\sigma^0-\sigma^0\partial_u\bar\sigma^0,
\end{equation}
and satisfy the evolution equations,
\begin{equation}
  \label{eq:75}
  \partial_u\Psi^0_3=\bar\partial\Psi^0_4,\quad \partial_u\Psi^0_2=\bar\partial\Psi^0_3+\sigma^0\Psi^0_4,\quad\partial_u\Psi^0_1=\bar\partial\Psi^0_2+2\sigma^0\Psi^0_3.
\end{equation}
We define here non-radiative spacetimes by the conditions
\begin{equation}
  \label{eq:80}
  \Psi^0_4=0=\Psi^0_3,
\end{equation}
and their complex conjugates. Furthermore, we require $\Psi^0_2$ to be real,
\begin{equation}
  \label{eq:83}
  \Psi^0_2=\bar\Psi^0_2.
\end{equation}
On account of the relations \eqref{eq:74}, these are constraints on the
asymptotic part of the shear and the news that are somewhat weaker than
$\partial_u \sigma^0=0$ and its complex conjugate, together with
$\partial^2\sigma^0=\bar\partial\bar\sigma^0$, i.e., the requirement that the
news vanishes together with the analog of the electric condition. The reason is
that the latter would require $\bar\partial^3 \bar Y=0=\partial^3Y$ and
eliminate superrotations, while the former are invariant under extended BMS$_4$
transformations. With these conditions, the map
\begin{equation}
  \label{eq:84}
  m(-\frac{1}{G}\Psi^0_2)= P_0^g,\quad m(-\frac{1}{2G}\Psi^0_1)=\bar J_0^g,\quad m(-\frac{1}{2G}\bar \Psi^0_1)= J_0^g
\end{equation}
is compatible with the transformations laws, while the remaining non-trivial
evolution equations \eqref{eq:75} are compatible with \eqref{eq:71}.

\section{Discussion}
\label{sec:discussion}

The Chern-Simons to chiral Wess-Zumino-Witten
\cite{Witten:1988hf,Moore:1989yh,Elitzur:1989nr} to Liouville theory
\cite{Forgacs:1989ac,Bershadsky:1989mf} reductions have been used in the context
of three-dimensional gravity
\cite{Coussaert:1995zp,Henneaux:1999ib,Barnich:2013yka} (see also
\cite{Cotler2019,Henneaux:2019sjx,Merbis2020}). They provide a direct approach
to constructing holographic action principles that may be compared to the group
theoretic constructions \cite{Barnich:2017jgw,Nguyen2021}.

Due to the absence of a pure Chern-Simons formulation, this avenue is less
straightforward in four dimensions, see however
\cite{deBoer:2003vf,AdamoCasaliSkinner2014,Nguyen:2020hot} for such
constructions. From this point of view, our models provide consistent targets
for holographically dual theories directly computed under suitable assumptions
from first order or second order gravitational action principles with surface
terms. Let us note however that the ones we have written here do not involve the
additional terms with Schwarzian derivatives due to central extensions of the
group that have played a prominent role in the three dimensions. In the case of
BMS$_3$ for instance, those terms are important to get the correct vacuum orbits
with a little group that is the Poincar\'e group in three dimensions
\cite{Barnich:2015uva}. In principle, such terms may also be added when
constructing geometric actions for the extended BMS$_4$ group with central
extensions. How Schwarzian derivatives or central extensions appear in the
gravitational context in four dimensions is discussed for instance in
\cite{nutku1992impulsive,Barnich:2016lyg,Strominger:2016wns,Barnich:2017ubf,Nguyen2022}.

More generally, considerations on the coadjoint orbits of closely related groups
that appear in gravitational theories may be found in
\cite{Donnelly2021,Ciambelli2022}, while prescriptions for gravitational charges
in terms of the left hand sides of \eqref{eq:84} have recently been discussed in
\cite{Compere:2020lrt,Freidel2022,Freidel2022a,Donnay2022}. Note that the
reality condition imposed in \eqref{eq:82} explicitly excludes magnetic mass,
which has played a prominent role in a number of recent studies
\cite{Henneaux:2004jw,Bunster:2006rt,Godazgar:2018qpq,Bunster:2019mup,Kol2019,Godazgar2020}.

In the case of the global BMS$_4$ group, the Maurer-Cartan forms for the
non-abelian part ${\rm SL}(2,\mathbb C)$ can be obtained directly from the
matrix representations
\begin{equation}
  \label{eq:82}
  g=\begin{pmatrix} a & b \\ c & d
    \end{pmatrix}=
    \begin{pmatrix}
      e^{-E/2} & - \bar A e^{E/2}\\ -B e^{-E/2} & (1+\bar A B) e^{E/2}
    \end{pmatrix}, 
\end{equation}
with $a,b,c,d,E,A,B\in \mathbb C$ and $ad-bc=1$. The second parametrization
corresponds to a composition of Lorentz rotations of type $II\circ I\circ II$ in
the terminology of \cite{Penrose:1984,Chandrasekhar:1985kt}. Explicit formulas
for the associated geometric action will be given elsewhere.

Finally, since BMS groups are conformal Carroll groups \cite{Duval:2014uva}, the
BMS$_3$ invariant models of \cite{Barnich:2012rz,Barnich:2017jgw} and the
BMS$_4$ invariant models constructed here are explicit examples of conformal
Carroll field theories in $1+1$ and $2+1$ dimensions, respectively.

\section*{Acknowledgments}
\label{sec:acknowledgements}

\addcontentsline{toc}{section}{Acknowledgments}

This work is supported by the F.R.S.-FNRS Belgium through convention FRFC PDR
T.1025.14 and convention IISN 4.4514.08. RR is supported by the Austrian Science
Fund (FWF), project P 32581-N. KN is grateful to Jakob Salzer for useful
discussions and supported by the STFC, grant numbers ST/P000258/1 and
ST/T000759/1.

\addcontentsline{toc}{section}{References}

\printbibliography

\end{document}